\documentstyle[12pt]{article}
\newcommand{\tr}[1]{\,{\rm tr}\,#1\,}

\begin{document}
\title{
\begin{flushright}
{\small SMI-15-97}
\end{flushright}
\vspace{0.5cm}
Phase of a chiral determinant and global $SU(2)$ anomaly.}
\author{A.A.Slavnov \thanks{E-mail:$~~$ slavnov@mi.ras.ru} \\Steclov 
Mathematical Institute, Russian Academy of Sciences, \\Gubkina st. 8,
GSP-1, 117966, Moscow, Russia} \maketitle

\begin{abstract}

A representation for the phase of a chiral determinant in terms of a 
path integral of a local action is constructed. This 
representation is used to modify the action of chiral $SU(2)$ fermions 
removing the global anomaly.
\end{abstract}

\section{Introduction} 

It is known that $SU(2)$ gauge models with an odd number of Weyl fermion
 doublets are affected by a global quantum anomaly leading to 
inconsistency of the theory \cite{W}. This anomaly is related to an 
ambiguity in the definition of the phase of a chiral determinant. It was
shown in the paper \cite{W} that one cannot define the phase of the
determinant of a single chiral $SU(2)$ fermion in a gauge invariant way. 

In the present paper I will get a representation for the phase of a chiral 
determinant as a path integral of exponent of a local action. 
Having this explicit representation one can modify the action of chiral 
$SU(2)$ fermions in such a way that the global anomaly disappears. My
construction in some sence is reminiscent to the Wess-Zumino construction 
for local quantum anomalies \cite{WZ}.

It also gives a representation for Atiyah-Patodi-Singer $ \eta$-invariant 
\cite{APS} as a path integral of a local action, as it was shown before
\cite{AG} that the phase of a chiral determinant is expressed in terms of 
$ \eta$.

\section{Lagrangian representation for the phase of 
a chiral determinant and global $SU(2)$ anomaly}

Let us start with the massless Euclidean Dirac operator
\begin{equation}
\hat{D}= \gamma_{\mu}( \partial_{\mu}+iA_{\mu}(x))
\label{1}
\end{equation}
which can be written in the form
\begin{equation}
 \hat{D}=\pmatrix{0&C \cr -C^+&0}
\label{2}
\end{equation}
where
\begin{equation}
 C=e_{\mu}(\partial_{\mu}+iA_{\mu}(x))
\label{3}
\end{equation}
with $e_i=-i \sigma_i, \sigma_i (i=1,2,3)$ being the Pauli matrices, $e_0=-I$.
The field $A_{\mu}(x)$ for each $x$ belongs to the Lie algebra of a gauge 
group.The matrices $C, C^+$ represent fermions of opposite chiralities.

It follows from the eq.(\ref{2}) that
\begin{equation}
\det{D}= \det{C} \det{C^+}
\label{4}
\end{equation}
Hence the determinant of the Dirac operator is equal to the square of the 
modulus of the determinant of the Weyl operator. In case of the $SU(2)$
 group
due to pseudoreality of representation 
\begin{equation}
 \det{C}= \det{C^+}
\label{5}
\end{equation} 
Using this fact Witten proposed to define the regularized determinant of
a single Weyl fermion as a square root of regularized Dirac determinant.
Dirac determinant can be regularized in a gauge invariant way by means
of a standard Pauli-Villars regularization. However there is a sign
ambiguity
\begin{equation}
 \det{C_R}= \pm (\det{D_R})^{1/2}
\label{6}
\end{equation}
For a given gauge field $A_{\mu}(x)$ one can choose a sign arbitrary. 
Defined in such a way the Weyl determinant is obviously invariant with 
respect to infinitesimal gauge transformations and therefore in the framework 
of perturbation theory one has a consistent gauge invariant definition of the
determinant of the chiral $SU(2)$ operator.

However if one allows topologically nontrivial gauge transformations, then 
one can choose a transformation $U$ which changes the sign of the square root 
of the determinant
\begin{equation}
[\det{D(A_{\mu})}]^{1/2}=- [\det{D(A_{\mu}^U)}]^{1/2}
\label{7}
\end{equation}
That means the phase of a chiral determinant is not invariant with respect
to "large" gauge transformations and leads to inconsistensy of the theory.

The definition of the chiral $SU(2)$ determinant as a square root of the 
regularized Dirac determinant is perfectly consistent in the framework of
perturbation theory, however it does not allow a representation of the Weyl
 determinant as a path integral of a local action as taking a square root of 
a determinant is a nonlocal operation. To get such a  representation for 
regularized chiral determinant we use the idea proposed in our paper \cite{SF}.
 
 We introduce the infinite set of Pauli-Villars fields with masses $Mr,
 \quad r=1,2 \ldots$. Now the regularized Weyl determinant may be
 written as follows
 \begin{equation}
 \det{C}_R= \int \exp \{ \int L_R dx \}d \bar{\psi}_+d \psi_+d \bar{\psi}_rd
 \psi_r
 \label{8}
 \end{equation}
 $$
 L_R= \bar{\psi}_+C \psi_+ + \sum_{r=1}^{ \infty} \bar{\psi}_r(
 \hat{D}+Mr) \psi_r
 $$
 Here $\psi_r$ are Pauli-Villars fields having Grassmanian parity
 $(-1)^{r+1}$.
 
 Integrating over $ \psi$ we get 
 \begin{equation}
 \det{C_R}= \det{C} \prod_{r=1}^{ \infty} \det{( \hat{D}+Mr)}^{(-1)^r}
 \label{9}
 \end{equation}
 Using the representation like eq.(\ref{2}) for the Dirac operator one
 can rewrite it as follows
 \begin{equation}
 \det{C_R}= \det{C} \prod_{r=1}^{\infty} \det{(|C|^2+M^2r^2)}^{(-1)^r}=
 \label{10}
 \end{equation}
 $$
 =\prod_{r=0}^{ \infty} \prod_iC_i^{-1}(|C_i|^2+M^2r^2)^{(-1)^r}
 $$
 Here $C_i$ are diagonal elements of the matrix $C_{ij}=<u_iCv_j>$,
 where $u_i$ and $v_j$ form orthonormal bases in the spaces of left
 and right handed fermions respectively.
 
 The product over $r$ can be calculated explicitely using the
 representation
 \begin{equation}
 \prod_{r=0}^{\infty}(|C_i|^2+M^2r^2)^{(-1)^r}= \exp \{
 \sum_{r=0}^{\infty} \ln(|C_i|^2+M^2r^2)(-1)^r \}
 \label{11}
 \end{equation}
 Differentiating the exponent with respect to $|C^i|^2$ one has
 \begin{equation}
 \frac{\partial}{\partial|C_i|^2} \sum_{r=0}^{\infty}(-1)^r
 \ln(|C_i|^2+M^2r^2) =1/2[\sum_{r=-
 \infty}^{\infty}(-1)^r(|C_i|^2+M^2r^2)^{-1}+|C_i|^{-2}]=
 \label{12}
 \end{equation}
 $$
 = 1/2[ \pi(M|C_i| \sinh( \pi|C_i|M^{-1}))^{-1}+|C_i|^{-2}]
 $$
 Integrating over $|C_i|^2$ one gets
 \begin{equation}
 \ln (\det{C_R})= \ln(\tan(\frac{\pi|C_i|}{2M}))
+ \ln(|C_i|)+\ln(B)
 \label{13}
 \end{equation}
where $B$ is a field independent constant which in the following is 
assumed to be included into normalization factor.
 Therefore up to normalization factor we have the following representation 
for the regularized determinant
 \begin{equation}
 \det{C_R}= \prod_i \frac{|C_i|}{C_i}
 \tan(\frac{\pi|C_i|}{2M})
 \label{14}
 \end{equation}
 In the framework of perturbation theory for a given $A_{\mu}$ one can
 fix the signs of $C_i$ at will, in particular take all $C_i$ positive.
 Then it follows from the eq.(\ref{14}) that
 \begin{equation}
 \det{C_R}= \prod_i \tan(\frac{\pi|C_i|}{2M}) 
 \label{15}
 \end{equation}
 One sees that all $|C_i|>>M$ are cutted and therefore eqs.(\ref{8},
 \ref{15})indeed provide the gauge invariant regularization of the
 chiral determinant. It can be demonstrated explicitely in terms of
 Feynmann diagrams. In particular for polarization operator one has
 \begin{equation}
 \Pi_{ij,R}^{\mu\nu}= \int \frac{d^4p}{(2 \pi)^4} \tr \{\frac{(1+
 \gamma_5)}{2} \frac{\gamma_{\mu} \tau_{i}(\hat{p}- \hat{k}) \gamma_{\nu} \tau_j
 \hat{p}}{(p-k)^2p^2} \}+
 \label{16}
 \end{equation}
 $$
 + \sum_{r=1}^{\infty} \int \frac{d^4p}{(2 \pi)^4} \tr \{
 \frac{\gamma_{\mu} \tau_i(\hat{p}- \hat{k}+Mr) \gamma_{\nu} \tau_j(
 \hat{p}+Mr)(-1)^r}{[(p-k)^2+M^2r^2][p^2+M^2r^2]} \}
 $$
 Using the standard technique one can rewrite it as follows
 \begin{equation}
 \Pi_{ij,R}^{\mu \nu}= \delta_{ij} \int \frac{d^4p}{(2 \pi)^4} \int_0^1
 d \alpha [1/2 \frac{ \tr[ \gamma_{\mu}( \hat{p}+ \alpha \hat{k}-
 \hat{k}) \gamma_{\nu}(\hat{p}+ \alpha \hat{k})}{[p^2+k^2( \alpha-
 \alpha^2)]^2}]+
 \label{17}
 \end{equation}
 $$
 + \sum_{r=1}^{\infty} \frac{ \tr[ \gamma_{\mu}( \hat{p}+ \alpha
 \hat{k}- \hat{k}+Mr) \gamma_{\nu}( \hat{p}+ \alpha
 \hat{k}+Mr)](-1)^r}{[p^2+k^2( \alpha- \alpha^2)+M^2r^2]^2}
 $$
 Separating the terms which do not contain $M$ in the nominator and
 performing the summation we get
 \begin{equation}
 \Pi_1 \sim \int \frac{d^4p}{(2 \pi)^4}1/2 \int_0^1d \alpha \tr[
 \gamma_{\mu}( \hat{p}+ \alpha \hat{k}- \hat{k}) \gamma_{\nu}( \hat{p}+
 \alpha \hat{k})] \times
\label{18}
\end{equation}
$$
 (- \frac{ \partial}{ \partial|p|^2} \pi(M \sqrt{p^2+k^2(
 \alpha- \alpha^2)} \sinh( \pi \sqrt{p^2+k^2( \alpha-
 \alpha^2)}M^{-1}))^{-1})
 $$
 One sees that the integrand decreases exponentially for large $p$
 providing fast convergence of the integral. The terms which contain $M$
 in the nominator are analyzed in the same way.

However beyond perturbation theory the expression (\ref{14}) is not well 
defined. Although large eigenvalues are cutted, the phase factor is not
 regularized. It is the source of the global anomaly in our approach. As 
was mentioned in the beginning, topologically nontrivial gauge transformations 
may change the sign of $C_i$ and one cannot fix it in arbitrary way. (A
different possibility to get the global anomaly starting from a local action
was discussed in ref. \cite{EN}, where embedding of the $SU(2)$ model into 
anomalous $SU(3)$ theory was considered).
 
 It is known that in the case of local anomalies one can restore gauge
 invariance by adding to the action a local term depending on new fields
 \cite{WZ}. Below I show that a similar construction is possible for the
 global anomaly as well.
 
 Let us modify the Lagrangian (\ref{7}) by adding to it a gauge
 invariant term describing the interaction of new fields $ \chi_r$
 $$
 L_R \rightarrow L'_R=L_R+ \Delta L
 $$
 \begin{equation}
 \Delta L= \bar{\chi}_+C \chi_+ + \sum_{r=1}^{\infty} \bar{\chi}_r(
 \hat{D}+m^2M^{-1}r) \chi_r
 \label{19}
 \end{equation}
 where $\chi_r$ are again the fields with alternating Grassmanian parity
 $(-1)^{r+1}$. Here $m$ is some fixed parameter with the dimension of
 mass. When $M \rightarrow \infty$ the masses of the $ \chi$ fields
 vanish.
 
 The integral over $ \chi$ can be calculated as above giving the result
 \begin{equation}
 \Delta= \prod_i \frac{|C_i|}{C_i} \tan( \frac{ \pi|C_i|}{2m^2M^{-1}})
 \label{20}
 \end{equation}
 Assuming that $C_i \neq 0$ we see that when $M \rightarrow \infty$
 \begin{equation}
 \Delta \rightarrow \prod_i \frac{|C_i|}{C_i}
 \label{21}
 \end{equation}
 It shows that the integral
 \begin{equation}
 \lim_{M \rightarrow \infty} \int \exp \{ \int \Delta L dx \}d \bar{\chi}_+ d
 \chi_+ d \bar{\chi}_rd \chi_r
 \label{22}
 \end{equation}
 gives the representation for the phase of a chiral determinant as a
 path integral of the local action. Note that this
 representation is valid for any gauge group, not necessary $SU(2)$.

One sees that the $\Delta$ exactly compensates the indefinite phase
 factor in the eq.(\ref{14}) and the integral
 \begin{equation}
 \det(C_R)'= \int \exp \{ \int L_R'dx \}d \bar{\psi}_+ d \psi_+ d \bar{\chi}_+ d \chi_+ d
 \bar{\psi}_r d \psi_r d \bar{ \chi}_r d \chi_r
 \label{23}
 \end{equation}
 provides a well defined expression which is gauge invariant not only
 with respect to infinitesimal gauge transformations but with respect to
 topologically nontrivial transformations as well.
 
 In the case of the $SU(2)$ gauge group the new fields $ \chi$ do not
 influence the results of perturbative calculations as in this case one
 can fix arbitrary the signs of $C_i$ and $ \lim_{M \rightarrow
 \infty}\Delta=1$. It is also seen from the explicit calculation of the
 polarization operator. When the mass of the $ \chi$ fields tends to
 zero the integral vanishes.
 
Analogous construction is valid for other models with global anomalies,
provided the fermions belong to a pseudoreal representation.

 \section{Discussion}
 
 We constructed above a representation for the phase of a chiral
 determinant in terms of a path integral of the exponent of the local
 action. Using this expression we were able to modify the
 action of $SU(2)$ chiral fermions in such a way that the global anomaly
 disappears. This construction has some similarity to the Wess-Zumino
 construction for local anomalies. There are however important
 differences. The Wess-Zumino term restores quantum gauge invariance, but
 the classical action including this term is not gauge invariant, where
 as our modified action (\ref{19}) is gauge invariant. Another
 difference is that contrary to the Wess-Zumino case, variation of our
 additional term under topologically nontrivial gauge transformation is
 discrete. The geometrical meaning of these terms is also different. As
 was mentioned above our construction gives a representation for
 Atiyach-Patodi-Singer $ \eta$-invariant.

 It would be interesting to analyze a possible physical meaning of the
 modified action. Although the new fields are not seen in perturbation
 theory, they certainly influence nonperturbative configurations. Even
 if one starts with the configuration which does not include the $ \chi$
 fields, they will be produced in pairs in a final state. One can
 speculate that the true vaccuum of the modified model includes infinite
 number of pairs of massless fermions which may change drastically a
 physical content of the theory.
 
 {\bf Acknowledgements.} \\
 This work was completed while the author was participating in the
 session of Benasque summer institute for theoretical physics. I am
 grateful to the organizers and in particular to A.Gonzales-Arroyo for 
 invitation and kind hospitality extended to me in Spain. I acknowledge 
useful discussions with L.Alvarez-Gaume and M.Asorey. This research is 
supported in part by Russian Basic Research Foundation under grant   
96-01-00551. $$~$$
 \end{document}